\documentclass[a4paper,11pt,usenames,dvipsnames]{article}

\usepackage{jhepsub}
\usepackage[T1]{fontenc}
\usepackage{amsmath}
\usepackage{slashed}
\usepackage{amssymb}
\usepackage[retainorgcmds]{IEEEtrantools}
\usepackage{simpler-wick}
\usepackage{tikz-feynman}
\usepackage{float}
\usepackage{verbatim}
\usepackage[bb=dsserif]{mathalpha}

\title{\color{Blue}\boldmath B-type anomaly coefficients for the \\ D3-D7 domain wall}
\author[a]{Georgios Linardopoulos}
\affiliation[a]{Asia Pacific Center for Theoretical Physics (APCTP) \\ Hogil Kim Memorial Building, \#501 POSTECH \\ 77 Cheongam-Ro Nam-gu, Pohang Gyeongsangbuk-do 37673, Korea.}
\emailAdd{georgios.linardopoulos@apctp.org}

\abstract{We compute B-type Weyl anomaly coefficients for the two non-supersymmetric codimension-1 domain wall versions of $\N = 4$ super Yang-Mills theory which are holographically dual to the D3-D7 probe-brane system. We compute the two-point function of the (improved) energy-momentum tensor in the presence of the domain wall, to leading order in perturbation theory. From it we determine the two-point function of the corresponding displacement operator and the $b_2$ anomaly coefficient for both domain wall systems.}

\arxivnumber{2502.13613}

\newcommand{\gym}{g_{\text{\scalebox{.8}{YM}}}}
\newcommand{\bL}{\textbf{L}}

\newcommand{\tr}{{\text{tr}}}
\newcommand{\x}{\textrm{x}}
\newcommand{\z}{\textrm{z}}
\newcommand{\kk}{\mathrm{k}}
\newcommand{\ii}{\mathfrak{i}}
\newcommand{\nn}{\mathfrak{n}}

\newcommand{\G}{\mathcal{G}}
\newcommand{\N}{\mathcal{N}}

\newcommand{\LL}{\mathcal{L}}
\newcommand{\I}{\mathcal{I}}
\newcommand{\D}{\mathcal{D}}
\newcommand{\Scal}{\mathcal{S}}

\newcommand{\quotes}[1]{``#1''}

\hypersetup{pdfstartview={XYZ null null 1}}
\allowdisplaybreaks
\begin{document} 
\begin{flushright}
APCTP Pre2025-002
\end{flushright}
\maketitle\flushbottom
\section[Introduction]{Introduction}
\noindent Even dimensional conformal field theories (CFTs) are known to develop conformal/Weyl (or trace) anomalies in curved spacetimes which make the trace of the energy-momentum (or stress) tensor acquire a non-vanishing expectation value. There is no Weyl anomaly in odd dimensions. Although flat space CFTs are generally free from Weyl anomalies, the corresponding (curved space) anomaly coefficients still show up in flat-space conformal data. For example, two and three-point functions of the energy-momentum tensor take the following forms in two spacetime dimensions $(x_1,x_2)$:
\begin{IEEEeqnarray}{c}
\left\langle T\left(\zeta_1\right\rangle T\left(\zeta_2\right)\right> = \frac{c/2}{\left(\zeta_1 - \zeta_2\right)^4}, \quad \left\langle T\left(\zeta_1\right\rangle T\left(\zeta_2\right) T\left(\zeta_3\right)\right> = \frac{c}{\left(\zeta_1 - \zeta_2\right)^2\left(\zeta_2 - \zeta_3\right)^2\left(\zeta_3 - \zeta_1\right)^2}, \qquad 
\end{IEEEeqnarray}
where $T \equiv T_{\zeta\zeta}$, and $\zeta \equiv x_1 + i x_2$, $\bar{\zeta} \equiv x_1 - i x_2$ are the holomorphic/anti-holomorphic coordinates, while $c$ is the single (ambient) Weyl anomaly coefficient there is in two dimensions. \\
\indent The presence of boundaries and defects gives rise to extra Weyl anomaly coefficients in both even and odd dimensions. The new anomaly coefficients are localized on the boundary. For example, the Weyl anomaly in the case of codimension-1 boundaries in 4-dimensional CFTs takes the following form (scheme-independent terms only) \cite{HerzogHuangJensen15, HerzogHuangJensen17} (see also \cite{ChalabiHerzogBannonRobinsonSisti21}):
\begin{IEEEeqnarray}{c}
\left\langle T^{\mu}_{\mu} \right\rangle^{d = 4} = \frac{1}{16\pi^2}\left(c \, W_{\mu\nu\rho\sigma}^2 - a \, E_4\right) + \frac{\delta\left(x_{\perp}\right)}{16\pi^2}\left(a \, E_4^{(\text{bry})} - b_1 \, \tr\hat{K}^3 - b_2 \, h^{pq}\hat{K}^{rs}W_{pqrs}\right), \qquad \label{WeylAnomaly4d1c}
\end{IEEEeqnarray}
where $W_{\mu\nu\rho\sigma}$ is the Weyl tensor, $E_4$ is the ambient Euler density, $E_4^{(\text{bry})}$ is the boundary term of the Euler characteristic, $K_{pq}$ is the boundary extrinsic curvature ($\hat{K}_{pq} \equiv K_{pq} - K h_{pq}/3$), and $h_{pq}$ is the induced metric on the boundary at $x_{\perp} = 0$. The Weyl anomaly \eqref{WeylAnomaly4d1c} includes two main types of ambient/boundary terms, the A-term $a$ and the B-terms $c$, $b_1$, $b_2$. \\
\indent Unlike 2d CFTs, where every correlation function of the energy-momentum tensor is determined by the single Weyl anomaly coefficient $c$, in four dimensions only two-point functions of the energy-momentum tensor are fully specified by the ambient B-type anomaly coefficient $c$. Three-point functions contain three independent unknown coefficients, so that only two of them can be fixed by the ambient anomaly coefficients $a$ and $c$ in \eqref{WeylAnomaly4d1c} \cite{OsbornPetkou93, ErdmengerOsborn96}. \\
\indent On the other hand, the (B-type) boundary anomaly coefficients $b_1$ and $b_2$ of codimension-1 defects in four dimensions fully determine the two and three point functions of the so-called displacement operator. These are given by:
\begin{IEEEeqnarray}{c}
\left\langle\D\left(\textbf{x}_1\right\rangle \D\left(\textbf{x}_2\right)\right> = \frac{c_{\eta\eta}}{\textbf{x}_{12}^{8}}, \qquad \left\langle\D\left(\textbf{x}_1\right) \D\left(\textbf{x}_2\right) \D\left(\textbf{x}_3\right)\right\rangle = \frac{c_{\eta\eta\eta}}{\textbf{x}_{12}^{4}\textbf{x}_{23}^{4}\textbf{x}_{31}^{4}}, \label{DisplacementOperatorTwoThreePointFunction}
\end{IEEEeqnarray}
where $\textbf{x}_{ij} \equiv \left|\textbf{x}_i - \textbf{x}_j\right|$ and the points $\textbf{x}_{i,j}$ lie on the 3d boundary.\footnote{We will be using roman typeface for the spacetime arguments of correlation functions in both CFTs and codimension-1 dCFTs, setting $\x = \left(\textbf{x},\z\right)$ to denote the presence of a codimension-1 boundary at $\z = 0$.} The displacement operator $\D\left(\textbf{x}\right)$ is a scalar operator that is localized on the codimension-1 boundary and quantifies the violation of translation invariance across it (see e.g.\ \cite{DrukkerMartelliShamir17} for a relevant discussion). The boundary anomaly coefficients $b_1$, $b_2$ are related to the two and three-point function structure constants of the displacement operator as follows \cite{HerzogHuangJensen17, HerzogHuang17}:
\begin{IEEEeqnarray}{c}
b_1 = \frac{2\pi^3}{35}\cdot c_{\eta\eta\eta}, \qquad b_2 = \frac{2\pi^4}{15}\cdot c_{\eta\eta}. \label{BtypeAnomalyCoefficients}
\end{IEEEeqnarray}
\noindent In free theories, the ambient B-type coefficient $c$ is related to the boundary B-type coefficient $b_2$ as $b_2 = 8c$ \cite{Fursaev15, Solodukhin15c}. This relation generally ceases to hold in interacting theories \cite{HerzogHuang17}. \\
\indent In the present paper we compute the B-type boundary anomaly coefficient $b_2$ for two 4-dimensional, codimension-1 defect CFTs (dCFTs) which are holographically dual to the D3-D7 probe-brane system. The first of these systems has an $SU(2)\times SU(2)$ global symmetry, while the second system has an $SO(5)$ global symmetry. Both dCFTs live in flat 4d space and have flat 3d defects/boundaries, so they are anomaly-free. Yet we can compute the corresponding B-type anomaly coefficients from the correlation functions of the energy-momentum tensor and the displacement operator, as we have just seen. Our computations are perturbative in the 't Hooft coupling constant $\lambda \equiv \gym^2 N_c$, and we report the leading-order result for the $b_2$ anomaly coefficients. Recently, a similar computation was carried out in \cite{deLeeuwKristjansenLinardopoulosVolk23} for the codimension-1 dCFT$_4$ which is holographically dual to the $SU(2)$ symmetric D3-probe-D5 brane system. \\
\indent One important difference between the two (codimension-1, 4-dimensional, flat-space) dCFTs is that the dCFT which is dual to the D3-D5 system is half-BPS supersymmetric, while both dCFTs which are dual to the D3-D7 system are non-supersymmetric. The first direct consequence of this is that the D3-D7 system is unstable and needs to be stabilized by appropriate worldvolume fluxes \cite{DavisKrausShah08, Rey08, MyersWapler08, BergmanJokelaLifschytzLippert10}. The second direct consequence concerns the integrability of the D3-D7 system, for which it is known that the $SU(2)\times SU(2)$ symmetric case is not integrable \cite{deLeeuwKristjansenVardinghus19}, while the $SO(5)$ symmetric case is only known to be integrable to leading order in perturbation theory \cite{deLeeuwKristjansenLinardopoulos18a, deLeeuwGomborKristjansenLinardopoulosPozsgay19, GomborBajnok20a} (see also \cite{Linardopoulos25a}). What is more, supersymmetric localization methods which have recently become available for the supersymmetric D3-D5 dCFT \cite{RobinsonUhlemann17, Robinson17, Wang20a, KomatsuWang20, BeccariaCaboBizet23, HeUhlemann25}, are not expected to be applicable to any of the non-supersymmetric dCFTs which are dual to the D3-D7 probe-brane system. \\
\indent Two-point functions of the energy-momentum tensor are very interesting observables in codimension-1 dCFTs \cite{McAvityOsborn93, McAvityOsborn95}. They are determined up to three (generally dependent) unknown coefficients which depend on the dCFT invariant ratio $v$, cf.\ \eqref{TwoPointFunctionDCFT}. Because one-point functions of the energy-momentum tensor vanish in codimension-1 dCFTs, two-point functions can be used together with the operator product expansion (OPE) to derive all higher-point correlation functions of the energy-momentum tensor. This is one of the main goals of the boundary conformal bootstrap program \cite{LiendoRastellivanRees12}, which aims to determine all defect CFT correlation functions from a minimal set of scalar and spinorial data (such as for example one and two-point function structure constants of the energy-momentum tensor). \\
\indent Our paper is organized as follows. In the following section \ref{Section:DomainWallD3D7}, we introduce the two domain wall setups which are holographically dual to the $SU(2)\times SU(2)$ and the $SO(5)$ symmetric D3-D7 probe-brane systems. In section \ref{Section:EnergyMomentumTensor} we compute the leading-order contribution to the energy-momentum tensor two-point function, for both versions of the D3-D7 domain wall. In section \ref{Section:AnomalyCoefficients} we compute the corresponding two-point functions of the displacement operator and from it we determine the B-type boundary anomaly coefficients of the D3-D7 domain wall system (both versions). Our conclusions can be found in section \ref{Section:Conclusions}.
\section[The D3-D7 domain wall]{The D3-D7 domain wall \label{Section:DomainWallD3D7}}
\noindent The D3-D7 probe-brane system is made up from a large number of $N_c \rightarrow \infty$ coincident D3 branes which intersect a single (probe) D7 brane. The near-horizon geometry of the D3-branes is AdS$_5\times \text{S}^5$. The relative orientation of the D-branes in this space is shown in table \ref{Table:D3D7system} below. The coordinates $x_0, \ldots, x_3, z$ parametrize AdS$_5$ in the Poincaré frame, while the angles $\psi, \theta, \chi, \vartheta, \varrho$ parametrize the 5-sphere S$^5$. 
\renewcommand{\arraystretch}{1.1}\setlength{\tabcolsep}{5pt}
\begin{table}[H]\begin{center}\begin{tabular}{|c||c|c|c|c|c|c|c|c|c|c|}
\hline
& ${\color{Red}x_0}$ & ${\color{Red}x_1}$ & ${\color{Red}x_2}$ & ${\color{Red}x_3}$ & ${\color{Red}z}$ & ${\color{Blue}\psi}$ & ${\color{Blue}\theta}$ & ${\color{Blue}\chi}$ & ${\color{Blue}\vartheta}$ & ${\color{Blue}\varrho}$ \\ \hline
\text{D3} & $\bullet$ & $\bullet$ & $\bullet$ & $\bullet$ &&&&&& \\ \hline
\text{D7} & $\bullet$ & $\bullet$ & $\bullet$ & & $\bullet$ & & $\bullet$ & $\bullet$ & $\bullet$ & $\bullet$ \\ \hline
\end{tabular}\caption{The D3-D7 intersection.\label{Table:D3D7system}}\end{center}\end{table}
\vspace{-.5cm}\indent It can be shown that the geometry of the probe D7-brane inside AdS$_5\times \text{S}^5$ can be either AdS$_4\times \text{S}^2\times \text{S}^2$ or AdS$_4\times \text{S}^4$ \cite{DavisKrausShah08, Rey08, MyersWapler08, BergmanJokelaLifschytzLippert10}. The respective global bosonic symmetries are therefore $SO(3,2)\times SO(3)\times SO(3)$ and $SO(3,2)\times SO(5)$, breaking all (maximal) supersymmetries of the D3-brane system. There is also a tachyonic instability which causes the compact parts of the D7-brane (S$^2\times\text{S}^2$ and S$^4$) to \quotes{slip off} either side of the S$^5$ equator. To lift the instability we add $k_1$ and $k_2$ units of magnetic flux on either of the two 2-spheres of the AdS$_4\times \text{S}^2\times \text{S}^2$ symmetric brane and $d_G \equiv (n+1)(n+2)(n+3)/6$ units of instanton flux on the 4-sphere of the AdS$_4\times \text{S}^4$ symmetric brane. For more information, see e.g.\ \cite{KristjansenSemenoffYoung12b, Linardopoulos25a}. \\
\indent The D3-D7 probe-brane system is holographically dual to a 4-dimensional, codimension-1 dCFT. The ambient CFT$_4$ is just $\N = 4$ SYM. Refer to appendix \ref{Appendix:SuperYangMills} for the Lagrangian and our basic conventions regarding $\N = 4$ SYM. At $x_3 = 0$ there is a flat $2+1$ dimensional boundary which may or may not host additional (mainly fermionic) degrees of freedom \cite{Rey08}. Defect CFT correlation functions can be computed at weak 't Hooft coupling $\lambda$ by means of \quotes{interfaces} or domain walls \cite{NagasakiTanidaYamaguchi11, NagasakiYamaguchi12}. These are described by classical (\quotes{fuzzy funnel}) solutions of the ambient equations of motion which share the global symmetries of the defect CFT \cite{ConstableMyersTafjord99, ConstableMyersTafjord01a}. \\
\indent For the domain wall which is dual to the $SU(2)\times SU(2)$ symmetric D3-probe-D7 brane system, the classical solution of the scalar equations of motion of $\N = 4$ SYM reads \cite{KristjansenSemenoffYoung12b}:
\begin{IEEEeqnarray}{l}
\varphi_{i} = \varphi_{i}^{\text{cl}}\left(x_3\right) = \frac{1}{x_3} \cdot \left[\begin{array}{cc} \big(t_i^{(k_1)} \otimes \mathbb{1}_{k_2}\big)_{k\times k} & 0_{k\times \left(N_c - k\right)} \\ 0_{\left(N_c - k\right)\times k} & 0_{\left(N_c - k\right)\times \left(N_c - k\right)} \end{array}\right]_{N_c\times N_c}, \quad i=1,2,3, \qquad \label{FuzzyFunnelSU2a} \\[6pt]
\varphi_{i} = \varphi_{i}^{\text{cl}}\left(x_3\right) = \frac{1}{x_3} \cdot \left[\begin{array}{cc} \big(\mathbb{1}_{k_1} \otimes t_i^{(k_2)}\big)_{k\times k} & 0_{k\times \left(N_c - k\right)} \\ 0_{\left(N_c - k\right)\times k} & 0_{\left(N_c - k\right)\times \left(N_c - k\right)} \end{array}\right]_{N_c\times N_c}, \quad i=4,5,6, \qquad \label{FuzzyFunnelSU2b}
\end{IEEEeqnarray}
where $x_3 >0$, and the fermionic/vector fields of $\N = 4$ SYM have been set to zero, i.e.\ $A_{\mu} = \psi_{\alpha} = 0$. For $k_1, k_2 = 0,1,\ldots$ we have also set,
\begin{IEEEeqnarray}{l}
k \equiv k_1 \cdot k_2,
\end{IEEEeqnarray}
while the matrices $t_i^{(k_{\ii})}$ furnish a $k_{\ii} \times k_{\ii}$ irreducible representation of $SU(2)$ (where $\ii = 1,2$):
\begin{IEEEeqnarray}{c}
\big[t_i^{(k_{\ii})}, t_j^{(k_{\ii})}\big] = i \epsilon_{ijl}t_l^{(k_{\ii})}, \qquad i,j,l = 1,2,3,
\end{IEEEeqnarray}
so that the domain wall solution \eqref{FuzzyFunnelSU2a}--\eqref{FuzzyFunnelSU2b} shares the global bosonic symmetry of its dual $SU(2)\times SU(2)$ symmetric D3-D7 probe-brane system. \\
\indent For the domain wall that is dual to the $SO(5)$ symmetric D3-D7 probe D-brane system, the classical solution of the scalar equations of motion of $\N = 4$ SYM is given by \cite{KristjansenSemenoffYoung12b}:
\begin{IEEEeqnarray}{l}
\varphi_{i} = \varphi_{i}^{\text{cl}}\left(x_3\right) = \frac{1}{\sqrt{8} x_3} \cdot \left[\begin{array}{cc} \left(G_i\right)_{d_G\times d_G} & 0_{d_G\times \left(N_c - d_G\right)} \\ 0_{\left(N_c - d_G\right)\times d_G} & 0_{\left(N_c - d_G\right)\times \left(N_c - d_G\right)} \end{array}\right]_{N_c\times N_c}, \quad i=1,\ldots,5, \qquad \label{FuzzyFunnelSO5a} \\
\varphi_{6} = 0, \label{FuzzyFunnelSO5b}
\end{IEEEeqnarray}
where $x_3 >0$, and the fermionic/vector fields of $\N = 4$ SYM have been set to zero as before ($A_{\mu} = \psi_{\alpha} = 0$). We have also defined,
\begin{IEEEeqnarray}{l}
d_G \equiv \frac{1}{6} \cdot (n+1)(n+2)(n+3), \qquad n = 1,2,\ldots \label{ConstantdG}
\end{IEEEeqnarray}
The five $d_G\times d_G$ fuzzy S$^4$ matrices ($G$-matrices) $G_i$ are given in terms of the following completely symmetrized (\quotes{sym}) tensor product \cite{CastelinoLeeTaylor97}:
\begin{IEEEeqnarray}{c}
G_i \equiv \left[\underset{n \text{ terms}}{\underbrace{\overset{n \text{ factors}}{\overbrace{\gamma_i\otimes\mathbb{1}_{4}\otimes\ldots\otimes\mathbb{1}_{4}}} + \mathbb{1}_{4}\otimes\gamma_i\otimes\ldots\otimes\mathbb{1}_{4} + \ldots + \mathbb{1}_{4}\otimes\ldots\otimes\mathbb{1}_{4}\otimes\gamma_i}}\right]_{\text{sym}}, \qquad \label{G-matrices}
\end{IEEEeqnarray}
where $i = 1,\ldots,5$ and $\gamma_i$ are the five $4\times4$ Euclidean Dirac matrices (in five dimensions):
\begin{IEEEeqnarray}{c}
\gamma_i = \left(\begin{array}{cc} 0 & -i\sigma_i \\ i\sigma_i & 0 \end{array}\right), \quad i = 1,2,3, \qquad \gamma_4 = \left(\begin{array}{cc} 0 & \mathbb{1}_2 \\ \mathbb{1}_2 & 0 \end{array}\right), \qquad \gamma_5 = \left(\begin{array}{cc} \mathbb{1}_2 & 0 \\ 0 & -\mathbb{1}_2 \end{array}\right),
\end{IEEEeqnarray}
while $\sigma_i$ are the three Pauli matrices. The Dirac matrices satisfy the $SO(5)$ Clifford algebra,
\begin{IEEEeqnarray}{lc}
\left\{\gamma_i,\gamma_j\right\} = 2\delta_{ij}\mathbb{1}_4.
\end{IEEEeqnarray}
The ten commutators of the five $G$-matrices,
\begin{IEEEeqnarray}{lc}
G_{ij} \equiv \frac{1}{2}\left[G_i,G_j\right], \label{Gcommutators}
\end{IEEEeqnarray}
furnish a $d_G$-dimensional (anti-hermitian) irreducible representation of $SO\left(5\right)\simeq Sp\left(4\right)$:
\begin{IEEEeqnarray}{lc}
\left[G_{ij},G_{kl}\right] = 2\left(\delta_{jk}G_{il} + \delta_{il}G_{jk} - \delta_{ik}G_{jl} - \delta_{jl}G_{ik}\right).
\end{IEEEeqnarray}
This way, the domain wall solution \eqref{FuzzyFunnelSO5a}--\eqref{FuzzyFunnelSO5b} shares the global bosonic symmetry of the $SO(5)$ symmetric D3-D7 system. More properties of the fuzzy S$^4$ matrices \eqref{G-matrices} can be found in references \cite{CastelinoLeeTaylor97, ConstableMyersTafjord01a}. \\
\indent To compute defect CFT correlation functions at weak coupling we assign vacuum expectation values (vevs) to the gauge fields. The vevs originate from the breaking of gauge symmetry across the defect; the values of the vevs are given by the fuzzy funnel solutions. By using the fuzzy funnel solutions \eqref{FuzzyFunnelSU2a}--\eqref{FuzzyFunnelSU2b} and \eqref{FuzzyFunnelSO5a}--\eqref{FuzzyFunnelSO5b}, tree-level one-point functions of chiral primary operators were computed in \cite{KristjansenSemenoffYoung12b} for both dCFTs which are dual to the D3-D7 probe-brane system. In the double scaling limit $\lambda/k^2, \lambda/n^2 \rightarrow 0$, the weak-coupling results so computed agreed with the corresponding supergravity calculation at strong coupling with Witten diagrams. \\
\indent The computation of tree-level one-point functions of non-protected scalar operators, which led eventually to the discovery of a closed-form determinant formula for all tree-level one-point functions of scalar operators in the $SO(5)$ symmetric D3-D7 dCFT, was carried out in \cite{deLeeuwKristjansenLinardopoulos16, deLeeuwGomborKristjansenLinardopoulosPozsgay19}. These works followed \cite{deLeeuwKristjansenZarembo15} which introduced the use of the Bethe ansatz for the computation of defect CFT correlation functions. The spectra of quantum fluctuations and one-loop corrections to the one-point functions of the vacuum state were worked out in \cite{GimenezGrauKristjansenVolkWilhelm18, GimenezGrauKristjansenVolkWilhelm19}. Agreement with the corresponding calculations at strong coupling was once more reported in the double scaling limit $\lambda/k^2, \lambda/n^2 \rightarrow 0$. Non-integrability of the $SU(2)\times SU(2)$ symmetric dCFT was shown in \cite{deLeeuwKristjansenVardinghus19}, while leading-order integrability of the $SO(5)$ symmetric dCFT was proven in \cite{deLeeuwKristjansenLinardopoulos18a}. The latter result was based on a (quench) integrability criterion which was formulated by Piroli, Pozsgay and Vernier in \cite{PiroliPozsgayVernier17} and followed from on the seminal work of Ghoshal and Zamolodchikov \cite{GhoshalZamolodchikov93}. In a follow-up paper \cite{GomborBajnok20a}, Gombor and Bajnok claimed that integrability of the $SO(5)$ symmetric setup breaks down beyond leading order. More information can be found in the reviews \cite{deLeeuw19, Linardopoulos20, KristjansenZarembo24a}.
\section[Energy-momentum tensor]{Energy-momentum tensor \label{Section:EnergyMomentumTensor}}
\noindent Having set up the domain wall descriptions of the two defect CFTs which are dual to the D3-D7 probe-brane system, we are now in position to compute correlation functions of the energy-momentum tensor at weak coupling. From these we can directly extract the correlation functions of the displacement operator and the corresponding boundary anomaly coefficients, as we will see in the next section. First off, let us go through the computation of ambient two and three-point functions.
\paragraph{Generalities} The form of two and three-point functions of the energy-momentum tensor in any $d$-dimensional CFT is completely fixed by symmetry, while one-point functions vanish \cite{OsbornPetkou93, ErdmengerOsborn96}. Two-point functions are in particular given by
\begin{IEEEeqnarray}{ll}
\left\langle\Theta_{\mu\nu}\left(\x_1\right) \Theta_{\rho\sigma}\left(\x_2\right)\right\rangle = \frac{C_{T}}{\x_{12}^{2d}} \cdot \I_{\mu\nu\rho\sigma}\left(\x_1 - \x_2\right), \qquad \x_{12} \equiv \left|x_1 - x_2\right|, \label{TwoPointFunctionCFT}
\end{IEEEeqnarray}
where the energy-momentum tensor has been improved (i.e.\ it is symmetric, conserved and traceless, cf.\ \eqref{EnergyMomentumTensorProperties}) and the inversion tensors $\I_{\mu\nu}$, $\I_{\mu\nu\rho\sigma}$ are defined as
\begin{IEEEeqnarray}{l}
\I_{\mu\nu}\left(\x\right) \equiv g_{\mu\nu} - \frac{2\,\x_{\mu}\x_{\nu}}{\x^2}, \quad \I_{\mu\nu\rho\sigma}\left(\x\right) \equiv \frac{1}{2}\left(\I_{\mu\rho}\left(\x\right)\I_{\nu\sigma}\left(\x\right) + \I_{\mu\sigma}\left(\x\right)\I_{\nu\rho}\left(\x\right)\right) - \frac{1}{d}\,g_{\mu\nu}g_{\rho\sigma}. \qquad \label{InversionTensors}
\end{IEEEeqnarray}
A similar (albeit more complicated) expression holds for three-point functions. In $\N = 4$ super Yang-Mills (SYM) theory, two and three-point functions of the energy-momentum tensor are protected \cite{AnselmiFreedmanGrisaruMarcusJohansen97, FreedmanMathurMatusisRastelli98a, HoweSokatchevWest98}. In other words, the leading-order values of the (improved) energy-momentum tensor two and three-point functions do not receive quantum corrections. As such they are both given by their free-field values (although it is quite straightforward to obtain the result by performing all the Wick contractions with the Feynman rules of $\N = 4$ SYM). For two-point functions in $d=4$ dimensions \cite{OsbornPetkou93}, 
\begin{IEEEeqnarray}{c}
C_T = \frac{N_0 + 3N_{1/2} + 12N_1}{3\pi^4}, \label{TwoPointFunctionStructureConstantsFree}
\end{IEEEeqnarray}
where $N_0$, $N_{1/2}$ and $N_1$ are respectively the numbers of free real scalar, spin one-half (Weyl/chiral basis), and spin-one fields. In the case of $\N = 4$ SYM (for which, $N_0 = 6N_c^2$, $N_{1/2} = 4N_c^2$, $N_1 = N_c^2$),
\begin{IEEEeqnarray}{c}
C_T = \frac{10N_c^2}{\pi^4}.
\end{IEEEeqnarray}
\indent In the presence of a codimension-1 boundary at $\z = 0$ (in $d$ spacetime dimensions), one-point functions of the (improved) energy-momentum tensor vanish,
\begin{IEEEeqnarray}{ll}
\left\langle\Theta_{\mu\nu}\left(\x_1\right)\right\rangle = 0,
\end{IEEEeqnarray}
while the form of two-point functions is once more fully specified by symmetry \cite{McAvityOsborn93, McAvityOsborn95}:
\begin{IEEEeqnarray}{l}
\left\langle\Theta_{\mu\nu}\left(\x_1\right)\Theta_{\rho\sigma}\left(\x_2\right)\right\rangle = \frac{1}{\x_{12}^{2d}} \cdot \Bigg\{\left(X_{\mu}X_{\nu} - \frac{g_{\mu\nu}}{d}\right)\left(X'_{\rho}X'_{\sigma} - \frac{g_{\rho\sigma}}{d}\right) A\left(v\right) + \Big(X_{\mu}X'_{\rho}I_{\nu\sigma} + \nonumber \\
+ X_{\mu}X'_{\sigma}I_{\nu\rho} + X_{\nu}X'_{\sigma}I_{\mu\rho} + X_{\nu}X'_{\rho}I_{\mu\sigma} - \frac{4}{d} \, g_{\mu\nu}X'_{\rho}X'_{\sigma} - \frac{4}{d} \, g_{\rho\sigma}X_{\mu}X_{\nu} + \frac{4}{d^2}\,g_{\mu\nu}g_{\rho\sigma}\Big) B\left(v\right) + \nonumber \\
\hspace{11cm} + \I_{\mu\nu\rho\sigma} C\left(v\right)\Bigg\}, \qquad \label{TwoPointFunctionDCFT}
\end{IEEEeqnarray}
however this time there is an explicit dependence on the dCFT invariant ratio $v$ which can be formed out of only 2 ambient points. The dCFT invariant ratio $v$, which enters the defect two-point function \eqref{TwoPointFunctionDCFT} through the functions $A\left(v\right)$, $B\left(v\right)$, $C\left(v\right)$, is defined as:
\begin{IEEEeqnarray}{c}
\xi \equiv \frac{\x_{12}^2}{4\left|\z_1\right|\left|\z_2\right|}, \qquad v^2 \equiv \frac{\xi}{\xi + 1} = \frac{\x_{12}^2}{\x_{12}^2 + 4\left|\z_1\right|\left|\z_2\right|}. \qquad \label{DefectInvariantRatios}
\end{IEEEeqnarray}
The definitions of the inversion tensors $\I_{\mu\nu}$ and $\I_{\mu\nu\rho\sigma}$ can be found in \eqref{InversionTensors} above. For simplicity, we have omitted their arguments $\left(\x_1 - \x_2\right)$ in \eqref{TwoPointFunctionDCFT}. We have also set,
\begin{IEEEeqnarray}{l}
X_{\mu} \equiv \z_1 \cdot \frac{v}{\xi} \frac{\partial\xi}{\partial \x_1^{\mu}} = v\left(\frac{2\z_1}{\x_{12}^2}\left(\x_{1\mu}-\x_{2\mu}\right) - \eta_{\mu}\right) \\
X'_{\rho} \equiv \z_2 \cdot \frac{v}{\xi} \frac{\partial\xi}{\partial \x_2^{\rho}} = -v\left(\frac{2\z_2}{x_{12}^2}\left(\x_{1\rho}-\x_{2\rho}\right) + \eta_{\rho}\right), \qquad
\end{IEEEeqnarray}
where $\eta \equiv \left(\textbf{0},1\right)$ is the unit normal to the defect at $\z = 0$. We also note that $X$, $X'$ obey,
\begin{IEEEeqnarray}{l}
X_{\mu}X_{\mu} = X_{\rho}'X_{\rho}' = 1, \qquad X_{\rho}' = \I_{\rho\mu}X_{\mu}.
\end{IEEEeqnarray}
The functions $A(v)$, $B(v)$ and $C(v)$ of the codimension-1 defect two-point function \eqref{TwoPointFunctionDCFT} are not completely independent, but instead they satisfy the following condition \cite{McAvityOsborn95}:
\begin{IEEEeqnarray}{l}
\left(v\,\frac{d}{dv} - d\right)\alpha\left(v\right) = 2(d-1)\gamma\left(v\right), \label{TwoPointFunctionCondition}
\end{IEEEeqnarray}
where the functions $\alpha(v)$ and $\gamma(v)$ are defined in terms of $A(v)$, $B(v)$ and $C(v)$ as follows:
\begin{IEEEeqnarray}{c}
\alpha(v) \equiv \frac{d-1}{d^2} \cdot \left[(d-1)\left(A(v) + 4B(v))+ d \, C(v)\right)\right], \qquad \gamma\left(v\right) \equiv -B(v) - \frac{C(v)}{2}. \qquad \label{TwoPointFunctionConditionVariables}
\end{IEEEeqnarray}
Far away from the boundary (where $\z_{1,2} \rightarrow \infty$ and $\xi, v \rightarrow 0$), conformal symmetry is restored and the dCFT two-point function \eqref{TwoPointFunctionDCFT} reduces to the CFT one in \eqref{TwoPointFunctionCFT}. Therefore, $A(0)= B(0) = 0$ and $C(0) = C_T$. Moreover, \eqref{TwoPointFunctionConditionVariables} tells us that $\alpha(0) = (d-1)C_T/d$.
\paragraph{Perturbative expansion} Let us now employ the domain wall description of the D3-D7 dCFT to calculate the two-point function of the energy-momentum tensor. Before being able to do that, we need to expand the fields of $\N = 4$ SYM theory \eqref{LagrangianSYM} around either of the classical solutions \eqref{FuzzyFunnelSU2a}--\eqref{FuzzyFunnelSU2b} and \eqref{FuzzyFunnelSO5a}--\eqref{FuzzyFunnelSO5b}:
\begin{IEEEeqnarray}{c}
A_{\mu} = \tilde{A}_{\mu}^{a}T^a, \quad \psi_{\alpha,m} = \tilde{\psi}_{\alpha,m}^{a}T^a, \quad \varphi_i\left(x\right) = \varphi_i^{\text{cl}}\left(x_3\right) + \tilde{\varphi}_i\left(x\right), \quad \tilde{\varphi}_i\left(x\right) = \tilde{\varphi}_i^a\left(x\right)T^a, \qquad \label{FieldPerturbationD3D5}
\end{IEEEeqnarray}
where $T^a$ are the ($N_c\times N_c$) generators of $U(N_c)$ in the fundamental representation. By inserting the perturbation \eqref{FieldPerturbationD3D5} into the expression of the improved energy-momentum tensor of $\N = 4$ SYM in \eqref{EnergyMomentumTensorImprovedSYM}, we are led to the following perturbative expansion:
\begin{IEEEeqnarray}{c}
\Theta_{\mu\nu}\left(x\right) = \sum_{n=0}^4 \Theta_{\mu\nu}^{(n)}\left(x\right) = \Theta_{\mu\nu}^{(0)} + \Theta_{\mu\nu}^{(1)} + \Theta_{\mu\nu}^{(2)} + \Theta_{\mu\nu}^{(3)}\left(x\right) + \Theta_{\mu\nu}^{(4)}, \label{EnergyMomentumTensorExpansion}
\end{IEEEeqnarray}
where the superscripts in the parentheses denote the number of perturbed fields. \\
\indent The first term $\Theta_{\mu\nu}^{(0)}$ in the perturbative expansion \eqref{EnergyMomentumTensorExpansion} can be obtained by simply plugging the fuzzy-funnel solutions \eqref{FuzzyFunnelSU2a}--\eqref{FuzzyFunnelSU2b}, \eqref{FuzzyFunnelSO5a}--\eqref{FuzzyFunnelSO5b} into the expression \eqref{EnergyMomentumTensorImprovedSYM} for the improved energy-momentum tensor of the ambient theory, that is $\N = 4$ SYM. Only the scalar part \eqref{EnergyMomentumTensorScalar} of the energy-momentum tensor \eqref{EnergyMomentumTensorImprovedSYM} is relevant, since the fermion and the vector boson fields have no vevs. We find that the classical value of the energy-momentum tensor vanishes,
\begin{IEEEeqnarray}{ll}
\Theta_{\mu\nu}^{\text{cl}} \equiv \Theta_{\mu\nu}^{(0)} = 0, \label{OnePointFunctionDefect}
\end{IEEEeqnarray}
for both the $SU(2)\times SU(2)$ symmetric domain wall \eqref{FuzzyFunnelSU2a}--\eqref{FuzzyFunnelSU2b}, and the $SO(5)$ symmetric one \eqref{FuzzyFunnelSO5a}--\eqref{FuzzyFunnelSO5b}. Obviously, so do the corresponding one-point functions, in full accordance with the analysis of McAvity-Osborn in \cite{McAvityOsborn93, McAvityOsborn95} and Liendo-Rastelli-van Rees in \cite{LiendoRastellivanRees12}. \\
\indent To compute the second term $\Theta_{\mu\nu}^{(1)}$ in the perturbative expansion \eqref{EnergyMomentumTensorExpansion}, we adopt the following unifying notation for the D3-D7 fuzzy funnel solutions \eqref{FuzzyFunnelSU2a}--\eqref{FuzzyFunnelSU2b} and \eqref{FuzzyFunnelSO5a}--\eqref{FuzzyFunnelSO5b}:
\begin{IEEEeqnarray}{l}
\varphi_{i} = \varphi_{i}^{\text{cl}}\left(x_3\right) = \frac{1}{x_3} \cdot \left[\begin{array}{cc} \left(\tau_i\right)_{\kk\times \kk} & 0_{\kk\times \left(N_c - \kk\right)} \\ 0_{\left(N_c - \kk\right)\times \kk} & 0_{\left(N_c - \kk\right)\times \left(N_c - \kk\right)} \end{array}\right]_{N_c\times N_c}, \quad i=1,\ldots,6. \label{GenericDomainWall}
\end{IEEEeqnarray}
We dub the upper-left, $\kk\times \kk$ dimensional diagonal block as the \quotes{massive} one, and the lower-right, $\left(N_c - \kk\right)\times \left(N_c - \kk\right)$ diagonal block as \quotes{massless}. The two remaining, $\kk\times \left(N_c - \kk\right)$ and $\left(N_c - \kk\right)\times \kk$ dimensional massless blocks of the $N_c \times N_c$ matrices \eqref{GenericDomainWall} will be called \quotes{off-diagonal}. The same nomenclature can also be applied to the perturbed fields:
\begin{IEEEeqnarray}{c}
\tilde{\varphi}_i = \begin{bmatrix}

    {\color{Blue}\begin{bmatrix}
    * & * \\ * & *
    \end{bmatrix}_{\kk \times \kk}} &
    \begin{matrix} * & * \\ * & * \end{matrix} \\

     \begin{matrix}
    * & * \\ * & *
    \end{matrix} &
    {\color{Red}\begin{bmatrix} * & * \\ * & * \end{bmatrix}}

    \end{bmatrix}_{N_c\times N_c}, \qquad \label{PerturbedFieldMatrices}
\end{IEEEeqnarray}
so that the fields in the \quotes{massive} block are Wick-contracted with one set of (massive) defect propagators, the fields in the off-diagonal blocks are Wick-contracted with another set of (massive) defect propagators, while the fields in the massless block are Wick-contracted with the massless propagators of $\N = 4$ SYM. \\
\indent The leading contribution to the perturbative expansion \eqref{EnergyMomentumTensorExpansion} takes the following form:
\begin{IEEEeqnarray}{ll}
\Theta_{\mu\nu}^{(1)}\left(x\right) = &\frac{1}{\gym^2} \, \frac{4}{3 x_3^2} \cdot \tr\bigg\{\Big(\frac{1}{x_3} \eta_{\mu} \eta_{\nu} \tilde{\varphi}_i + \eta_{\mu} \partial_{\nu}\tilde{\varphi}_i + \eta_{\nu} \partial_{\mu}\tilde{\varphi}_i - \frac{g_{\mu\nu}}{2} \, \partial_{3}\tilde{\varphi}_i + \frac{x_3}{2} \, \partial_{\mu}\partial_{\nu}\tilde{\varphi}_i\Big) \cdot \tau_i + \nonumber \\[6pt]
&+ \frac{g_{\mu\nu}}{2x_3} \left(2\tau_j \tau_i \tau_j - \tau_i \tau_j \tau_j - \tau_j \tau_j \tau_i \right) \cdot \tilde{\varphi}_i \bigg\}. \label{EnergyMomentumTensorLeading}
\end{IEEEeqnarray}
For the $SU(2)\times SU(2)$ symmetric D3-D7 domain wall \eqref{FuzzyFunnelSU2a}--\eqref{FuzzyFunnelSU2b}, $\tau_i = t_i^{(k_1)} \otimes \mathbb{1}_{k_2}$ and $\tau_{i+3} = \mathbb{1}_{k_1} \otimes t_i^{(k_2)}$ (for $i=1,2,3$) in \eqref{GenericDomainWall}, so that the leading correction \eqref{EnergyMomentumTensorLeading} becomes:
\begin{IEEEeqnarray}{ll}
\Theta_{\mu\nu}^{(1)}\left(x\right) = \frac{1}{\gym^2} \, \frac{4}{3 x_3^2} \cdot \tr\bigg\{\Big(\frac{1}{x_3} \left(\eta_{\mu} \eta_{\nu} - g_{\mu\nu} \right) \tilde{\varphi}_i + \eta_{\mu} \partial_{\nu}\tilde{\varphi}_i &+ \eta_{\nu} \partial_{\mu}\tilde{\varphi}_i - \frac{g_{\mu\nu}}{2} \, \partial_{3}\tilde{\varphi}_i + \nonumber \\
& + \frac{x_3}{2} \, \partial_{\mu}\partial_{\nu}\tilde{\varphi}_i\Big) \cdot \tau_i\bigg\}, \qquad \label{EnergyMomentumTensorLeadingSU2}
\end{IEEEeqnarray}
where we have used
\begin{IEEEeqnarray}{ll}
\tau_i \tau_i = (c_1 + c_2)\cdot\mathbb{1}_k, \qquad \tau_j \tau_i \tau_j = \left(c_1 + c_2 - 1\right)\tau_i, \qquad c_1 \equiv \frac{k_1^2 - 1}{4}, \quad c_2 \equiv \frac{k_2^2 - 1}{4}. \qquad \label{ConstantsSU2}
\end{IEEEeqnarray}
For the $SO(5)$ symmetric D3-D7 domain wall \eqref{FuzzyFunnelSO5a}--\eqref{FuzzyFunnelSO5b}, $\tau_i = G_i/\sqrt{8}$ ($i = 1,\ldots,5$) and the leading correction \eqref{EnergyMomentumTensorLeading} in this case reads:
\begin{IEEEeqnarray}{ll}
\Theta_{\mu\nu}^{(1)}\left(x\right) = \frac{1}{\gym^2} \, \frac{4}{3 x_3^2} \cdot \tr\bigg\{\Big(\frac{1}{x_3} \left(\eta_{\mu} \eta_{\nu} - g_{\mu\nu} \right) \tilde{\varphi}_i + \eta_{\mu} \partial_{\nu}\tilde{\varphi}_i &+ \eta_{\nu} \partial_{\mu}\tilde{\varphi}_i - \frac{g_{\mu\nu}}{2} \, \partial_{3}\tilde{\varphi}_i + \nonumber \\
& + \frac{x_3}{2} \, \partial_{\mu}\partial_{\nu}\tilde{\varphi}_i\Big) \cdot \frac{G_i}{\sqrt{8}}\bigg\}, \qquad \label{EnergyMomentumTensorLeadingSO5}
\end{IEEEeqnarray}
where we have used the following properties of $G$-matrices \cite{CastelinoLeeTaylor97, ConstableMyersTafjord01a} in \eqref{EnergyMomentumTensorLeading}:
\begin{IEEEeqnarray}{ll}
G_i G_i = n(n+4) \cdot \mathbb{1}_{d_G}, \qquad G_j G_i G_j = \left(n(n+4) - 8\right)G_i.
\end{IEEEeqnarray}
Because all the terms of the leading corrections $\Theta_{\mu\nu}^{(1)}$ in \eqref{EnergyMomentumTensorLeadingSU2} and \eqref{EnergyMomentumTensorLeadingSO5} are essentially traces of products of unperturbed fields like \eqref{GenericDomainWall} and perturbed fields like \eqref{PerturbedFieldMatrices}, it is only the fields in the \quotes{massive} $\kk\times\kk$ blocks which contribute to $\Theta_{\mu\nu}^{(1)}$.
\paragraph{Two-point function} Once we have at our disposal a perturbative framework \eqref{EnergyMomentumTensorExpansion} for the computation of the energy-momentum tensor in the dCFT that is dual to the D3-D7 probe-brane system, we can start computing correlation functions. To get the $\nn$-point connected correlation function, we multiply $\nn$ copies of the (improved) energy-momentum tensor \eqref{EnergyMomentumTensorExpansion} (evaluated at different ambient points, $\x_1, \ldots,\x_{\nn}$) and Wick-contract all the perturbed fields with the defect propagators. We proceed order-by-order in the number of perturbed fields, starting with a single Wick contraction (since $\Theta_{\mu\nu}^{(0)} = 0$) and continuing until all four perturbed fields $\tilde{\varphi}_i$ (which can be present in the $\Theta_{\mu\nu}^{(4)}$ term of the energy-momentum tensor \eqref{EnergyMomentumTensorImprovedSYM}) are Wick-contracted. \\
\indent Since each entry of the energy-momentum tensor contributes a factor of $1/\lambda$ and each Wick contraction is proportional to $\lambda$, the leading-order term of the $\nn$-point function will be proportional to $\lambda^{1-\nn}$, while each subsequent term will get an extra Wick contraction and an additional factor of $\lambda$. For the (connected) two-point function, the perturbative expansion takes the following form:\footnote{We ignore lollipop diagram contributions which have been shown to vanish in the dCFT which is dual to the D3-D5 probe-brane system \cite{Buhl-MortensenLeeuwIpsenKristjansenWilhelm16c}.}
\begin{IEEEeqnarray}{c}
\left\langle\Theta_{\mu\nu}\left(\x_1\right)\Theta_{\rho\sigma}\left(\x_2\right)\right\rangle = \scriptsize{\feynmandiagram [small, inline =(a.base), horizontal= a to b]
{a [dot] --[blue, edge label=\(\color{black}\lambda^{-1}\)] b [dot]};} + \scriptsize{\feynmandiagram [small, inline =(a.base), horizontal= a to b]
{a [dot] -- [blue, out=45, in=135, edge label=\(\color{black}\lambda^0\)] b [dot] -- [blue, out=225, in=315] a};} +
\scriptsize{\feynmandiagram [small, inline =(a.base), horizontal= a to b]
{a [dot] -- [blue, edge label=\(\qquad\color{black}\lambda^0\)] b [dot], b -- [blue, out=315, in=45, loop, min distance=1cm] b};} + \scriptsize{\feynmandiagram [small, inline =(a.base), horizontal= a to b]
{a [dot] -- [blue, out=45, in=135, edge label=\(\color{black}\lambda\)] b [dot] -- [blue, out=225, in=315] a, a -- [blue] b};} + \scriptsize{\feynmandiagram [small, inline =(a.base), horizontal= a to b]
{a [dot] -- [blue, out=45, in=135, edge label=\(\qquad \color{black}\lambda\)] b [dot] -- [blue, out=225, in=315] a, b -- [blue, out=315, in=45, loop, min distance=1cm] b};} +
\scriptsize{\feynmandiagram [small, inline =(a.base), horizontal= a to b]
{a [dot] -- [blue, out=45, in=135] b [dot] -- [blue, out=225, in=315] a, a -- [blue, half left, edge label=\(\color{black}\lambda^2\)] b -- [blue, half left] a};} \qquad \quad
\label{EnergyMomentumTensorTwoPointFunction}
\end{IEEEeqnarray}
\indent The leading contribution to the connected part of the energy-momentum tensor two-point function $\langle \Theta_{\mu\nu}^{(1)}\Theta_{\rho\sigma}^{(1)}\rangle$, consists of a single Wick contraction and it is of order $\lambda^{-1}$. Based on what we discussed above, only the scalar fields in the massive $\kk\times\kk$ block contribute. The corresponding propagators for both versions of the D3-D7 domain wall can be found in appendix \ref{Appendix:Propagators} (cf.\ \eqref{PropagatorSU2mixed} and \eqref{PropagatorSU2pure1}--\eqref{PropagatorSU2pure2} for the $SU(2) \times SU(2)$ symmetric case, \eqref{PropagatorSO5complicated} and \eqref{PropagatorSO5easy} for $SO(5)$). Using these, we compute the following Wick-contracted quantities:
\begin{IEEEeqnarray}{c}
\wick{\tr[\tau_i \c1{\tilde{\varphi}_i}] \cdot \tr[\tau_j \c1{\tilde{\varphi}_j}]} = k\left(c_1 + c_2\right) K^{5/2}\left(x,y\right) = k\left(c_1 + c_2\right) \frac{\gym^2}{320\pi^2} \frac{1}{x_3 y_3} \frac{{}_2 F_1\left(2,3,6 ; -\xi^{-1}\right)}{\xi^3\left(1 + \xi\right)} \qquad \ \label{WickContraction1} \\
\wick{\tr[G_i \c1{\tilde{\varphi}_i}] \cdot \tr[G_j \c1{\tilde{\varphi}_j}]} = c_n \cdot K^{5/2}\left(x,y\right) = c_n \cdot \frac{\gym^2}{320\pi^2} \frac{1}{x_3 y_3} \frac{{}_2 F_1\left(2,3,6 ; -\xi^{-1}\right)}{\xi^3\left(1 + \xi\right)}, \label{WickContraction2}
\end{IEEEeqnarray}
where we have defined,
\begin{IEEEeqnarray}{ll}
c_n \equiv \tr\left[G_i G_i\right] = \frac{1}{6}\cdot n(n+1)(n+2)(n+3)(n+4), \quad n = 1,2,\ldots. \label{ConstantSO5}
\end{IEEEeqnarray}
To compute the leading-order contribution to the energy-momentum tensor two-point function, we only need the leading-order terms $\Theta_{\mu\nu}^{(1)}$ of the $\Theta_{\mu\nu}$ perturbative expansion \eqref{EnergyMomentumTensorExpansion}. These are given by \eqref{EnergyMomentumTensorLeadingSU2} for the $SU(2)\times SU(2)$ symmetric domain wall and by \eqref{EnergyMomentumTensorLeadingSO5} for the $SO(5)$ symmetric domain wall. Multiplying out two from each of these terms and performing all the Wick contractions we arrive at (for both domain wall systems):
\begin{IEEEeqnarray}{ll}
\big\langle\Theta_{\mu\nu}^{(1)}\left(\x_1\right)\Theta_{\rho\sigma}^{(1)}\left(\x_2\right)\big\rangle = \frac{1}{\x_{12}^8} \Bigg\{\left(X_{\mu}X_{\nu} - \frac{g_{\mu\nu}}{4}\right)\left(Y_{\rho}Y_{\sigma} - \frac{g_{\rho\sigma}}{4}\right) A\left(v\right) + \Big(X_{\mu}Y_{\rho}I_{\nu\sigma} + X_{\mu}Y_{\sigma}I_{\nu\rho} + \nonumber \\
+ X_{\nu}Y_{\sigma}I_{\mu\rho} + X_{\nu}Y_{\rho}I_{\mu\sigma} - g_{\mu\nu}Y_{\rho}Y_{\sigma} - g_{\rho\sigma}X_{\mu}X_{\nu} + \frac{1}{4}\,g_{\mu\nu}g_{\rho\sigma}\Big) B\left(v\right) + \I_{\mu\nu\rho\sigma} C\left(v\right)\Bigg\}. \qquad \label{EnergyMomentumTensorTwoPointFunctionLeading}
\end{IEEEeqnarray}
The result \eqref{EnergyMomentumTensorTwoPointFunctionLeading} we find, maintains the generic form of the energy-momentum tensor two-point functions \eqref{TwoPointFunctionDCFT} which was specified in \cite{McAvityOsborn93, McAvityOsborn95}, with
\begin{IEEEeqnarray}{l}
\I_{\mu\nu\rho\sigma}\left(\x\right) \equiv \frac{1}{2}\left(\I_{\mu\rho}\left(\x\right)\I_{\nu\sigma}\left(\x\right) + \I_{\mu\sigma}\left(\x\right)\I_{\nu\rho}\left(\x\right)\right) - \frac{1}{4}\,g_{\mu\nu}g_{\rho\sigma}.
\end{IEEEeqnarray}
In both versions of the D3-D7 domain wall system, the functions of the dCFT invariant ratio $A(v)$, $B(v)$ and $C(v)$ are given by
\begin{IEEEeqnarray}{l}
A\left(v\right) = 4 \gamma_{\kk} \left(6v^6 + 3v^4 + v^2\right) \label{ConstantA} \\
B\left(v\right) = - \gamma_{\kk} \left(3v^6 - v^4 - 2v^2\right) \label{ConstantB} \\
C\left(v\right) = \gamma_{\kk} v^2 \left(v^2 - 1\right)^2, \qquad \label{ConstantC}
\end{IEEEeqnarray}
where
\begin{IEEEeqnarray}{c}
\gamma_{\kk} \equiv \frac{32 c_{\kk}}{9 \pi^2 \gym^2}, \qquad c_{\kk} = \left\{\begin{array}{ll} k\left(c_1 + c_2\right) \qquad & SU(2)\times SU(2) \\[6pt] c_n/8 \qquad & SO(5), \end{array}\right.
\end{IEEEeqnarray}
while $k \equiv k_1k_2$, and the $SU(2)\times SU(2)$ constants $c_1$ and $c_2$ were defined in \eqref{ConstantsSU2}. See \eqref{ConstantSO5} for the definition of the $SO(5)$ constant $c_n$. As in the case of the D3-D5 domain wall which was recently studied in \cite{deLeeuwKristjansenLinardopoulosVolk23}, the functions $A(v)$, $B(v)$ and $C(v)$ in \eqref{ConstantA}--\eqref{ConstantC} satisfy the condition \eqref{TwoPointFunctionCondition}.
\section[Anomaly coefficients]{Anomaly coefficients \label{Section:AnomalyCoefficients}}
\noindent In the present section we will compute the B-type boundary anomaly coefficient $b_2$ for both versions of the dCFT which is dual to the D3-D7 probe-brane system. As we have already seen in \eqref{DisplacementOperatorTwoThreePointFunction}--\eqref{BtypeAnomalyCoefficients}, these anomaly coefficients are directly related to the two-point function structure constants of the displacement operator. In turn, the displacement operator is related to the energy-momentum tensor in a simple way as we will see below, so that we can use the results we have found above for the two-point function of the energy-momentum tensor in order to extract the corresponding two-point function of the displacement operator and its structure constants. \\
\indent Let us first briefly go through ambient anomalies. It has long been known that the ambient Weyl anomaly in four dimensions (given by the ambient terms of \eqref{WeylAnomaly4d1c}) is one-loop exact for theories with $\N = 4$ superconformal symmetry \cite{HoweSokatchevWest98}. As a result, the corresponding (ambient) anomaly coefficients $a$ and $c$ are given by their free-field values, just like the structure constants $C_T$ of the energy-momentum two-point functions in \eqref{TwoPointFunctionStructureConstantsFree}. The exact expressions for the ambient anomaly coefficients are \cite{BirrellDavies99}:
\begin{IEEEeqnarray}{c}
c = \frac{N_0 + 3N_{1/2} + 12N_1}{120} = \frac{\pi^4 C_T}{40}, \qquad a = \frac{2N_0 + 11N_{1/2} + 124N_1}{720}. \label{AnomalyCoefficientsAmbient}
\end{IEEEeqnarray}
In the case of $\N = 4$ SYM ($N_0 = 6N_c^2$, $N_{1/2} = 4N_c^2$, $N_1 = N_c^2$), the ambient anomaly coefficients are given by,
\begin{IEEEeqnarray}{c}
a = c = \frac{N_c^2}{4} = \frac{\pi^4 C_T}{40}. \label{AnomalyCoefficientsN=4sYM}
\end{IEEEeqnarray}
\indent We will now specify the boundary, B-type anomaly coefficient $b_2$, to leading order in the 't Hooft coupling constant $\lambda$, for both variants of the D3-D7 domain wall. Following \cite{HerzogHuangJensen17, HerzogHuang17}, we will read off the coefficient from the two-point function of the displacement operator, cf.\ \eqref{DisplacementOperatorTwoThreePointFunction}--\eqref{BtypeAnomalyCoefficients}. But first let us define the displacement operator. The displacement operator $\D$ is a scalar operator which is localized on the boundary/defect at $x_3 = 0$ and quantifies the breaking of translation invariance across it, as we have already mentioned. It is defined by the divergence of the improved energy-momentum tensor as follows:
\begin{IEEEeqnarray}{ll}
\partial^{\mu}\Theta_{\mu\nu}\left(x\right) = \delta(x_3) \,\eta_{\nu}\,\D\left(\textbf{x}\right), \label{DisplacementOperatorDefinition}
\end{IEEEeqnarray}
where $\eta_{\mu}$ is the unit normal to the boundary at $x_3 = 0$, and $x = \left(\textbf{x},x_3\right)$. Integrating the normal coordinate $x_3$ from $0^-$ to $0^+$, by also taking into account the conformal invariance of the defect, we find
\begin{IEEEeqnarray}{ll}
\D\left(\textbf{x}\right) = \lim_{x_3\rightarrow0+}\Theta_{33}\left(\textbf{x},x_3\right)-\lim_{x_3\rightarrow0-}\Theta_{33}\left(\textbf{x},x_3\right), \label{DisplacementOperatorEMtensor}
\quad
\end{IEEEeqnarray}
where $\textbf{x} \equiv \left(x_0,x_1,x_2\right)$. The leading-order contribution to the displacement operator two-point function follows directly from \eqref{EnergyMomentumTensorTwoPointFunctionLeading}, \eqref{DisplacementOperatorEMtensor}:
\begin{IEEEeqnarray}{c}
\big\langle\D^{(1)}\left(\textbf{x}_1\right)\D^{(1)}\left(\textbf{x}_2\right)\big\rangle = \lim_{\z_1,\z_2\rightarrow0^+}\big\langle\Theta^{(1)}_{33}\left(\x_1\right)\Theta^{(1)}_{33}\left(\x_2\right)\big\rangle
= \frac{c_{\eta\eta}}{\textbf{x}_{12}^{8}},
\end{IEEEeqnarray}
where the structure constant $c_{\eta\eta}$ is given by
\begin{IEEEeqnarray}{c}
c_{\eta\eta} \equiv \frac{15\,b_2}{2\pi^4} = \frac{80 \, c_{\kk} \, N_c}{\pi^2\lambda}.
\end{IEEEeqnarray}
Therefore the boundary B-type anomaly coefficient $b_2$ takes the following values for each of the two versions of the D3-D7 domain wall, to leading order in perturbation theory:
\begin{IEEEeqnarray}{c}
b_2 = \frac{32\pi^2 c_{\kk} N_c}{3 \lambda} + \mathcal{O}\left(\lambda^0\right), \qquad c_{\kk} = \left\{\begin{array}{ll} k\left(c_1 + c_2\right) \qquad & SU(2)\times SU(2) \\[6pt] c_n/8 \qquad & SO(5), \end{array}\right. \label{AnomalyCoefficientB2}
\end{IEEEeqnarray}
where again $k \equiv k_1k_2$, the D3-D7 domain wall constants $c_1$, $c_2$ were defined in \eqref{ConstantsSU2}, and $c_n$ in \eqref{ConstantSO5}.
\section[Conclusions]{Conclusions \label{Section:Conclusions}}
We have computed the B-type boundary anomaly coefficient $b_2$, to leading order in weak 't Hooft coupling $\lambda$, for both versions of the (non-supersymmetric) defect CFT that is dual to the D3-D7 probe-brane system. The dCFTs are described by domain walls; one domain wall has a global $SU(2)\times SU(2)$ symmetry and the other one has an $SO(5)$ global symmetry. \\
\indent Our result for the anomaly coefficient is \eqref{AnomalyCoefficientB2}. The 4-dimensional B-type anomaly coefficients $c$ and $b_2$ satisfy,
\begin{IEEEeqnarray}{c}
c = \frac{\pi^4}{30} \cdot \alpha(0), \qquad b_2 = \frac{2\pi^4}{15} \cdot \alpha(1), \label{BtypeAnomalyCoefficientsRelations}
\end{IEEEeqnarray}
where $\alpha\left(v\right)$ was defined in \eqref{TwoPointFunctionConditionVariables} above. The relation for the 4-dimensional B-type ambient anomaly coefficient $c$ is a direct consequence of $c = \pi^4 C_T/40$ in \eqref{AnomalyCoefficientsAmbient} and the identity $\alpha(0) = (d-1)C_T/d$ which we showed in section \ref{Section:EnergyMomentumTensor}. The relation for the 4-dimensional boundary anomaly coefficient $b_2$ follows from the identification \eqref{BtypeAnomalyCoefficients}, property \eqref{DisplacementOperatorEMtensor} of the displacement operator and the generic form \eqref{TwoPointFunctionDCFT} of energy-momentum tensor two-point functions in codimension-1 dCFTs (see e.g.\ \cite{McAvityOsborn95}). \\
\indent By using the values \eqref{ConstantA}--\eqref{ConstantC} for the functions $A(v)$, $B(v)$ and $C(v)$ that we found for the D3-D7 domain wall we can verify that $b_2$ in \eqref{AnomalyCoefficientB2} does indeed satisfy \eqref{BtypeAnomalyCoefficientsRelations}. For the ambient anomaly coefficient we find $c = 0$ which is also to be expected, given that our leading-order result \eqref{EnergyMomentumTensorTwoPointFunctionLeading} for the energy-momentum tensor two-point function consists of a single Wick contraction. On the other hand, to obtain the full result \eqref{AnomalyCoefficientsN=4sYM} for the ambient anomaly coefficient $c$, two Wick contractions between the fields of $\N = 4$ SYM are needed. As we have mentioned in the introduction, in free dCFTs the ambient and boundary B-type anomaly coefficients $c$ and $b_2$ are not independent as they satisfy $b_2 = 8c$. This identity is clearly not valid in our fully interacting setup, since $b_2 \sim \lambda^{-1} > 8c = 0$ (at least as long as $k_{1,2} > 1$), to leading order in the coupling constant $\lambda \rightarrow 0$. \\
\indent Working out higher-order perturbative corrections to the $b_2$ boundary anomaly coefficient would allow us to check whether $b_2 = 8c$ remains true when $k = 0$ or $k_1 = k_2 = 1$ (besides being an interesting computation in itself). In other words, we would be able to find out whether the boundary B-type anomaly coefficient $b_2$ is also one-loop exact (for $k = 0$ and $k_1 = k_2 = 1$) like the corresponding ambient anomaly coefficients $c$ and $a$. In the same vein, it would be interesting to calculate the B-type boundary anomaly coefficient $b_1$ in \eqref{BtypeAnomalyCoefficients}. This calculation is significantly harder of course, since now the three-point function of the energy-momentum tensor and the displacement operator are required. \\
\indent The AdS/{\color{Red} d}CFT correspondence \cite{NagasakiTanidaYamaguchi11, NagasakiYamaguchi12} opens the possibility for a wide range of computations of boundary anomaly coefficients in holographic defect CFTs. Holographic defects of varying codimensions can be studied, both at weak \cite{deLeeuwKristjansenLinardopoulosVolk23} and strong coupling (following e.g.\ the recent works \cite{GeorgiouLinardopoulosZoakos23, Linardopoulos25a}). For supersymmetric defects, localization methods \cite{RobinsonUhlemann17, Robinson17, Wang20a, KomatsuWang20} are also available. The program of computing new boundary anomaly coefficients in the context of the AdS/{\color{Red}d}CFT correspondence will be described in more detail elsewhere \cite{Linardopoulos25c}.
\newpage\acknowledgments
\noindent The author is thankful to M.\ de Leeuw, C.\ Kristjansen, P.\ Panopoulos, I.\ Papadimitriou, M.\ Roberts, M.\ Volk, and K.\ Zarembo for discussions. This work was supported by the National Development Research and Innovation Office (NKFIH) research grant K134946. The work of G.L.\ was supported in part by the National Research Foundation of Korea (NRF) grant funded by the Korea government (MSIT) (No.\ 2023R1A2C1006975), as well as by an appointment to the JRG Program at the APCTP through the Science and Technology Promotion Fund and Lottery Fund of the Korean Government.
\appendix 
\section[$\N = 4$ super Yang-Mills]{$\N = 4$ super Yang-Mills \label{Appendix:SuperYangMills}}
\paragraph{Lagrangian} The Lagrangian density of $\N = 4$ super Yang-Mills (SYM) theory reads:
\begin{IEEEeqnarray}{ll}
\LL_{\N = 4} = \frac{2}{\gym^2} \cdot \tr\bigg\{-\frac{1}{4} F_{\mu\nu} F^{\mu\nu} &- \frac{1}{2} \left(D_{\mu}\varphi_i\right)^2 + i\,\bar{\psi}_{\alpha}\slashed{D}\psi_{\alpha} + \frac{1}{4}\left[\varphi_i,\varphi_j\right]^2 + \nonumber \\
& + \sum_{i = 1}^{3}\G^i_{\alpha\beta}\bar{\psi}_{\alpha}\left[\varphi_i,\psi_{\beta}\right] + \sum_{i = 4}^{6}\G^i_{\alpha\beta} \bar{\psi}_{\alpha} \gamma_5 \left[\varphi_i,\psi_{\beta}\right]\bigg\}, \qquad \label{LagrangianSYM}
\end{IEEEeqnarray}
where $\bar{\psi}_{\alpha} \equiv \psi_{\alpha}^{\dagger} \gamma^0$, $\slashed{D} \equiv \gamma^{\mu}D_{\mu}$. In the present work, we adopt a mostly-plus $(-+++)$ signature convention for the Minkowski metric. The definitions of the four $4\times 4$ Minkowski Dirac matrices $\gamma^{\mu}$ in four dimensions (Weyl/chiral basis) and the six $4\times 4$ matrices $\G^{i}$, which enter the $\N = 4$ SYM Lagrangian density \eqref{LagrangianSYM}, can be found in many places, see e.g.\ the appendixes of \cite{Linardopoulos25a}. The fields of the Lagrangian \eqref{LagrangianSYM} carry adjoint $U(N_c)$ color indices for which,
\begin{IEEEeqnarray}{l}
A_{\mu} = A_{\mu}^{a}T^a, \qquad \varphi_{i} = \varphi_{i}^{a}T^a, \qquad \psi_{\alpha,m} = \psi_{\alpha,m}^{a}T^a, \qquad a = 1,\ldots, N_c^2, \label{ColorDecompositionSYM}
\end{IEEEeqnarray}
where $\mu = 0,\ldots,3$, $i = 1,\ldots,6$, $\alpha = 1,\ldots,4$ and $m = 1, \ldots, 4$. The equations of motion that follow from the Lagrangian \eqref{LagrangianSYM} are:
\begin{IEEEeqnarray}{l}
D^{\mu}F_{\mu\nu} = i\left[D_{\nu}\varphi_i,\varphi_i\right], \qquad D^{\mu}D_{\mu}\varphi_i = \left[\varphi_j,\left[\varphi_j,\varphi_i\right]\right], \\
i\slashed{D}\psi_{\alpha} = \sum_{i=1}^{3}\G_{\alpha\beta}^{i}\left[\psi_{\beta},\varphi_i\right] + \sum_{i=4}^{6} \G_{\alpha\beta}^{i}\gamma_5\left[\psi_{\beta},\varphi_i\right]. \qquad
\end{IEEEeqnarray}
\paragraph{Energy-momentum tensor} To compute the energy-momentum tensor of $\N = 4$ SYM, we may employ either the canonical prescription,
\begin{IEEEeqnarray}{ll}
T_{\mu\nu} = \frac{\partial\LL}{\partial\partial^{\mu}A_{\rho}}\,\partial_{\nu}A_{\rho} + \frac{\partial\LL}{\partial\partial^{\mu}\varphi_{i}}\,\partial_{\nu}\varphi_{i} + \frac{\partial\LL}{\partial\partial^{\mu}\bar{\psi}_{\alpha}}\,\partial_{\nu}\bar{\psi}_{\alpha} + \frac{\partial\LL}{\partial\partial^{\mu}\psi_{\alpha}}\,\partial_{\nu}\psi_{\alpha} - g_{\mu\nu}\LL, \label{EnergyMomentumTensorCanonical}
\end{IEEEeqnarray}
or the covariant prescription (see e.g.\ \cite{BirrellDavies99}),
\begin{IEEEeqnarray}{ll}
T_{\mu\nu} \equiv \frac{2}{\sqrt{-g}}\cdot\frac{\delta \Scal}{\delta g^{\mu\nu}}, \qquad \Scal = \int dx^4 \sqrt{-g} \, \LL, \label{EnergyMomentumTensorCovariant}
\end{IEEEeqnarray}
which leads to a manifestly symmetric energy-momentum tensor. On the other hand, the canonical energy-momentum tensor \eqref{EnergyMomentumTensorCanonical} is neither symmetric, traceless, or conserved and it needs to be improved \cite{CallanColemanJackiw70}. By applying a series of transformations to the canonical recipe \eqref{EnergyMomentumTensorCanonical} we are led to
\begin{IEEEeqnarray}{ll}
\Theta_{\mu\nu} = \frac{2}{\gym^2} \cdot \tr\bigg\{&-{F_{\mu}}^{\varrho} F_{\nu\varrho} - \frac{2}{3} \left(D_{\mu}\varphi_i\right) \left(D_{\nu}\varphi_i\right) + \frac{1}{3} \, \varphi_i \, D_{(\mu} D_{\nu)}\varphi_i + \frac{i}{2} \, \bar{\psi}_{\alpha}\gamma_{(\mu}\overset{\leftrightarrow}{D}_{\nu)}\psi_{\alpha}\bigg\} - \nonumber \\
& - \frac{2}{\gym^2} \cdot \tr\left\{-\frac{1}{4} \, F_{\mu\nu} F^{\mu\nu} - \frac{1}{6} \left(D_{\mu}\varphi_i\right)^2 - \frac{1}{12}\left[\varphi_i,\varphi_j\right]^2\right\} \cdot g_{\mu\nu}, \qquad \label{EnergyMomentumTensorImprovedSYM}
\end{IEEEeqnarray}
where $a_{(\mu\nu)} \equiv \left(a_{\mu\nu} + a_{\nu\mu}\right)/2$, and $f\overset{\leftrightarrow}{\partial}_{\mu} g \equiv f \left(\partial_{\mu}g\right) - \left(\partial_{\mu}f\right) g$. Details of the improvement procedure can be found in the appendixes of \cite{deLeeuwKristjansenLinardopoulosVolk23} and in many other places, such as e.g.\ the set of lectures \cite{Osborn19}. The improved energy-momentum tensor so constructed \eqref{EnergyMomentumTensorImprovedSYM} is manifestly symmetric, on-shell traceless and on-shell conserved:
\begin{IEEEeqnarray}{ll}
\Theta_{\mu\nu} = \Theta_{\nu\mu}, \qquad g^{\mu\nu} \Theta_{\mu\nu} = 0, \qquad \partial^{\mu}\Theta_{\mu\nu} = 0. \label{EnergyMomentumTensorProperties}
\end{IEEEeqnarray}
In the present work we will be mainly interested in the scalar part of improved energy-momentum tensor which is obtained from \eqref{EnergyMomentumTensorImprovedSYM} by setting the fermions and the vector bosons to zero:
\begin{IEEEeqnarray}{l}
\Theta_{\mu\nu}^{\text{scalar}} = \frac{2}{\gym^2} \tr\left\{-\frac{2}{3}\left(\partial_{\mu}\varphi_i\right)\left(\partial_{\nu}\varphi_i\right) + \frac{1}{3}\,\varphi_i\left(\partial_{\mu}\partial_{\nu}\varphi_i\right) + \frac{1}{6} \, g_{\mu\nu}\left[\left(\partial_{\varrho}\varphi_i\right)^2 + \frac{1}{2}\left[\varphi_i,\varphi_j\right]^2\right]\right\}. \qquad \label{EnergyMomentumTensorScalar}
\end{IEEEeqnarray}
The scalar part of the improved energy-momentum tensor \eqref{EnergyMomentumTensorScalar} is obviously symmetric, on-shell traceless and conserved.
\section[Scalar propagators in the D3-D7 dCFT]{Scalar propagators in the D3-D7 dCFT \label{Appendix:Propagators}}
\noindent The present appendix includes the expressions of all the scalar propagators that we use in this paper, for both versions of the defect CFT which is holographically dual to the D3-D7 probe brane system. As we have explained in section \ref{Section:EnergyMomentumTensor}, we only need the propagators in the \quotes{massive} blocks. More details, as well as derivations of the results included here can be found in the original papers \cite{GimenezGrauKristjansenVolkWilhelm18, GimenezGrauKristjansenVolkWilhelm19}.
\paragraph{$SU(2)_{k_1}\times SU(2)_{k_2}$ symmetric dCFT} We start off with the $SU(2)_{k_1}\times SU(2)_{k_2}$ symmetric domain wall \cite{GimenezGrauKristjansenVolkWilhelm18}. The fluctuations of the scalar fields (in the \quotes{massive} $k\times k$ block) are decomposed in fuzzy $SU(2)$ spherical harmonics as follows:
\begin{IEEEeqnarray}{l}
\left[\tilde{\varphi}_i\right]_{n_1,n_2} = \sum_{\ell_1 = 0}^{k_1 - 1} \sum_{\ell_2 = 0}^{k_2 - 1} \sum_{m_1 = - \ell_1}^{\ell_1} \sum_{m_2 = - \ell_2}^{\ell_2} \left(\tilde{\varphi}_i\right)_{\ell_1,m_1; \ell_2, m_2} \cdot \left[\hat{Y}_{\ell_1}^{m_1} \otimes \hat{Y}_{\ell_2}^{m_2}\right]_{n_1,n_2},
\end{IEEEeqnarray}
where $i = 1,\ldots,6$ and $n_1,n_2 = 1,\ldots, k \equiv k_1k_2$. All in all, there are three different kinds of scalar propagators, (namely $\langle\tilde{\varphi}_i \tilde{\varphi}_{j+3}\rangle$, $\langle\tilde{\varphi}_i \tilde{\varphi}_j\rangle$, $\langle\tilde{\varphi}_{i+3} \tilde{\varphi}_{j+3}\rangle$, for $i,j = 1,2,3$), depending on which branch of the $SU(2)\times SU(2)$ symmetric domain wall (i.e.\ \eqref{FuzzyFunnelSU2a} or \eqref{FuzzyFunnelSU2b}) we are. In the mixed sector we find,
\begin{IEEEeqnarray}{l}
{\color{Red}\langle(\tilde{\varphi}_i)_{\ell_1 m_1; \ell_2 m_2} (\tilde{\varphi}_{j+3})_{\ell_1' m_1'; \ell_2' m_2'}\rangle = (-1)^{m_1' + m_2'}\delta_{\ell_1\ell_1'}\delta_{\ell_2\ell_2'} [t^{(\ell_1)}_i]_{m_1,-m_1'} [t^{(\ell_2)}_{j+3}]_{m_2,-m_2'} K^{\varphi}_{\text{opp}}}, \qquad \label{PropagatorSU2mixed}
\end{IEEEeqnarray}
where the function $K^{\varphi}_{\text{opp}}$ is defined as:
\begin{IEEEeqnarray}{ll}
K^{\varphi}_{\text{opp}} \equiv \frac{K^{m^2_-}}{N_-} - \frac{K^{m_0^2}}{N_0} + \frac{K^{m^2_+}}{N_+}, \qquad &N_\pm \equiv \lambda_{\mp}\left(\lambda_{\mp} - \lambda_{\pm}\right), \quad N_0 \equiv -\lambda_+ \lambda_- \qquad \\
& \lambda_{\pm} \equiv -\frac{1}{2} \pm \sqrt{\ell_1 (\ell_1 + 1) + \ell_2 (\ell_2 + 1) + \frac{1}{4}}, \qquad
\end{IEEEeqnarray}
and the masses are given by,
\begin{IEEEeqnarray}{l}
m^2_0 \equiv \ell_1(\ell_1 + 1) + \ell_2(\ell_2 + 1) + 2, \qquad m^2_{\pm} \equiv \ell_1(\ell_1 + 1) + \ell_2(\ell_2 + 1) - 2\lambda_{\pm}.
\end{IEEEeqnarray}
See \eqref{ScalarPropagatorAdS3} below for the functional form of the propagators $K^{m_{0,\pm}^2}$. In the two pure sectors the following formulas hold:
\begin{IEEEeqnarray}{ll}
{\color{Red}\langle(\tilde{\varphi}_i)_{\ell_1 m_1; \ell_2 m_2} (\tilde{\varphi}_j)_{\ell_1' m_1'; \ell_2' m_2'}\rangle =}& \hspace{-.7cm} {\color{Red}(-1)^{m_1' + m_2'}\delta_{\ell_1\ell_1'}\delta_{\ell_2\ell_2'}\delta_{m_2 + m_2'} \bigg[\delta_{ij} \delta_{m_1 + m_1'} K^{\varphi,\ell_1}_{\text{sing}} -} \nonumber \\
&\hspace{-.7cm} {\color{Red}- i\epsilon_{ijk} [t^{(\ell_1)}_k]_{m_1, -m_1'} K^{\varphi,\ell_1}_{\text{anti}} - [t^{(\ell_1)}_i t^{(\ell_1)}_j]_{m_1, -m_1'} K^{\varphi,\ell_1}_{\text{sym}}\bigg]} \qquad \label{PropagatorSU2pure1} \\[6pt]
{\color{Red}\langle(\tilde{\varphi}_{i+3})_{\ell_1 m_1; \ell_2 m_2} (\tilde{\varphi}_{j+3})_{\ell_1' m_1'; \ell_2' m_2'}\rangle =} & {\color{Red}(-1)^{m_1' + m_2'}\delta_{\ell_1\ell_1'}\delta_{\ell_2\ell_2'}\delta_{m_1 + m_1'} \bigg[\delta_{ij} \delta_{m_2 + m_2'} K^{\varphi,\ell_2}_{\text{sing}} -} \nonumber \\
& {\color{Red}- i\epsilon_{ijk} [t^{(\ell_2)}_k]_{m_2, -m_2'} K^{\varphi,\ell_2}_{\text{anti}} - [t^{(\ell_2)}_i t^{(\ell_2)}_j]_{m_2, -m_2'} K^{\varphi,\ell_2}_{\text{sym}}\bigg]}, \label{PropagatorSU2pure2} \qquad
\end{IEEEeqnarray}
where the functions $K^{\varphi,\ell_i}_{\text{sing}}$, $K^{\varphi,\ell_i}_{\text{anti}}$ and $K^{\varphi,\ell_i}_{\text{sym}}$ are defined as follows, for $i = 1,2$:
\begin{IEEEeqnarray}{ll}
K^{\varphi,\ell_i}_{\text{sing}} \equiv \frac{\ell_i + 1}{2 \ell_i + 1} \cdot K^{m^2_{i,+}} + \frac{\ell_i}{2\ell_i + 1} \cdot K^{m^2_{i,-}}, \qquad K^{\varphi,\ell_i}_{\text{anti}} \equiv \frac{K^{m^2_{i,+}}}{2\ell_i + 1} - \frac{K^{m^2_{i,-}}}{2\ell_i + 1} \qquad \\[6pt]
K^{\varphi,\ell_1}_{\text{sym}} \equiv \frac{K^{m^2_{1,+}}}{(2 \ell_1 + 1)(\ell_1 + 1)} + \frac{K^{m^2_{1,-}}}{(2 \ell_1 + 1) \ell_1} - \frac{\ell_2 (\ell_2 + 1)}{\ell_1 (\ell_1 + 1)}\cdot\frac{K^{m^2_0}}{N_0} - \frac{K^{m^2_-}}{N_-} - \frac{K^{m^2_+}}{N_+} \\[6pt]
K^{\varphi,\ell_2}_{\text{sym}} \equiv \frac{K^{m^2_{2,+}}}{(2 \ell_2 + 1)(\ell_2 + 1)} + \frac{K^{m^2_{2,-}}}{(2 \ell_2 + 1) \ell_2} - \frac{\ell_1 (\ell_1 + 1)}{\ell_2 (\ell_2 + 1)}\cdot\frac{K^{m^2_0}}{N_0} - \frac{K^{m^2_-}}{N_-} - \frac{K^{m^2_+}}{N_+},
\end{IEEEeqnarray}
and the masses are given by:
\begin{IEEEeqnarray}{ll}
m^{2}_{1,+} = \ell_1 (\ell_1 - 1) + \ell_2 (\ell_2 + 1), \qquad & m^{2}_{1,-} = (\ell_1 + 1)(\ell_1 + 2) + \ell_2 (\ell_2 + 1) \qquad \\
m^{2}_{2,+} = \ell_1 (\ell_1 + 1) + \ell_2 (\ell_2 - 1), \qquad & m^{2}_{2,-} = \ell_1 (\ell_1 + 1) + (\ell_2 + 1)(\ell_2 + 2). \qquad 
\end{IEEEeqnarray}
The propagator functions $K^{m_{0,\pm}^2}$ and $K^{m_{i,\pm}^2}$ can again be found in \eqref{ScalarPropagatorAdS3} below. \\
\indent For the matrix elements of the $(2\ell+1)\times(2\ell+1)$ dimensional representation of the $SU(2)$ generators $t_i^{(2\ell+1)}$, we are using the following shorthand notation:
\begin{IEEEeqnarray}{l}
[t^{(\ell)}_i]_{m,-m'} \equiv [t^{(2\ell+1)}_i]_{\ell-m+1,\ell+m'+1}, \qquad i = 1,2,3,
\end{IEEEeqnarray}
where the $k\times k$ generators are given in terms of the fuzzy $SU(2)$ spherical harmonics $\hat{Y}^{m}_{\ell}$ by the following formulae:
\begin{IEEEeqnarray}{l}
t_1^{(k)} = \frac{\left(-1\right)^{k+1}}{2}\sqrt{\frac{k\left(k^2 - 1\right)}{6}} \cdot \left(\hat{Y}^{-1}_{1} - \hat{Y}^{1}_{1}\right) \label{t1FuzzySphericalHarmonics} \\[6pt]
t_2^{(k)} = \frac{i\left(-1\right)^{k+1}}{2}\sqrt{\frac{k\left(k^2 - 1\right)}{6}} \cdot \left(\hat{Y}^{-1}_{1} + \hat{Y}^{1}_{1}\right) \label{t2FuzzySphericalHarmonics} \\[6pt]
t_3^{(k)} = \frac{\left(-1\right)^{k+1}}{2}\sqrt{\frac{k\left(k^2 - 1\right)}{3}} \cdot \hat{Y}^{0}_{1}. \label{t3FuzzySphericalHarmonics}
\end{IEEEeqnarray}
Moreover, the fuzzy $SU(2)$ spherical harmonics satisfy the following identities:
\begin{IEEEeqnarray}{l}
\big(\hat{Y}^m_{\ell}\big)^\dagger = (-1)^m \hat{Y}^{-m}_{\ell} \qquad \& \qquad \tr\big[\hat{Y}^m_{\ell} \hat{Y}^{m'}_{\ell'}\big] = (-1)^{m'} \delta_{\ell\ell'} \delta_{m + m'}.
\end{IEEEeqnarray}
\paragraph{$SO(5)_{n}$ symmetric dCFT} In the case of the $SO(5)_n$ symmetric domain wall, we decompose the scalar fields of the \quotes{massive} $d_G \times d_G$ block in fuzzy $SO(5)$ spherical harmonics as follows \cite{GimenezGrauKristjansenVolkWilhelm19}:
\begin{IEEEeqnarray}{l}
\left[\tilde{\varphi}_i\right]_{n_1,n_2} = \sum_{\bL} \left(\tilde{\varphi}_i\right)_{\bL} \cdot \big[\hat{Y}_{\bL}\big]_{n_1,n_2}, \qquad \tr\left(\hat Y_{\bL'}^{\dagger} \hat{Y}_{\bL\phantom{'}}^{\phantom{\dagger}}\right) = \delta_{\bL', \bL\phantom{'}},
\end{IEEEeqnarray}
where $i = 1,\ldots, 5$ and $n_1,n_2 = 1,\ldots, d_G \equiv (n+1)(n+2)(n+3)/6$. We use the bold symbol $\bL$ to denote all the quantum numbers which specify states in representations of $SO(5)$. In particular, $\bL = \{(L_1,L_2), \ell_1, \ell_2, m_1, m_2\}$, where the quantum numbers $L_1$ and $L_2$ label the representation and the orbital/magnetic quantum numbers $\ell_i$ and $m_i$ obey,
\begin{IEEEeqnarray}{c}
-L_1 + L_2 \leq \ell_1 - \ell_2 \leq L_1 - L_2 \leq \ell_1 + \ell_2 \leq L_1 + L_2 \\
 \ell_1+\ell_2 \in \mathbb{Z}, \qquad m_i = -\ell_i,\ldots,\ell_i.
\end{IEEEeqnarray}
\indent Based on the description \eqref{FuzzyFunnelSO5a}--\eqref{FuzzyFunnelSO5b} of the $SO(5)$ symmetric D3-D7 domain wall, we obtain two kinds of scalar propagators in the \quotes{massive} block, namely $\langle\tilde{\varphi}_i \tilde{\varphi}_{j}\rangle$ (\quotes{complicated} fields, $i = 1,\ldots,5$), and $\langle\tilde{\varphi}_6 \tilde{\varphi}_6\rangle$ (\quotes{easy} fields). For the complicated scalars we have,
\begin{IEEEeqnarray}{ll}
{\color{Red}\langle(\tilde{\varphi}_i)_{\bL\phantom{'}}^{\phantom{\dagger}} (\tilde{\varphi}_j)_{\bL'}^\dagger\rangle =} & {\color{Red}\delta_{ij} \delta_{\bL, \bL'} \, \hat f^{\,\text{sing}} + \langle\bL|L_{ij}|\bL'\rangle \, \hat f^{\,\text{lin}} + \langle\bL|\{L_{ik}, L_{jl}\}L_{kl}|\bL'\rangle \, \hat f^{\,\text{cubic}} + } \qquad \nonumber \\
& {\color{Red} + \langle\bL|\{L_{ik},L_{kj}\}|\bL'\rangle \, \hat f^{\,\text{sym}}_{5} + \langle\bL|\{L_{i6}, L_{6j}\}|\bL'\rangle \cdot \Big[\delta_{L_1, L_1'} \delta_{L_2, L_2'} \, \hat f^{\,\text{sym}}_{6} +} \qquad \nonumber \\
& {\color{Red} + \delta_{L_1', L_1 \pm 1} \delta_{L_2', L_2 \mp 1} \, \hat f^{\,\text{opp}}\Big]}, \label{PropagatorSO5complicated}
\end{IEEEeqnarray}
where the expressions of the propagator functions $f^{\,\text{sing}}$, $f^{\,\text{lin}}$, $f^{\,\text{cubic}}$, $f^{\,\text{sym}}_{5}$, $f^{\,\text{sym}}_{6}$, and $f^{\,\text{opp}}$ can be found in appendix C of \cite{GimenezGrauKristjansenVolkWilhelm19}. In \eqref{PropagatorSO5complicated} we have defined,
\begin{IEEEeqnarray}{ll}
\langle\bL| L_{ij} |\bL'\rangle = \tr\Big(\hat{Y}_\bL^\dagger L_{ij}\hat{Y}_{\bL'}^{\phantom{\dagger}}\Big) = \frac{i}{2} \cdot \tr\Big(\hat{Y}_\bL^\dagger\big[\hat{Y}_{\bL'}^{\phantom{\dagger}}, G_{ij}\big]\Big), \qquad G_{i6} = -G_{6i} \equiv G_i,
\end{IEEEeqnarray}
for $i,j,k = 1,\ldots, 5$ and the commutators $G_{ij}$ of the $G$-matrices were defined in \eqref{Gcommutators} above. Also, $L_{ij} = - L_{ji}$ are the usual generators of $SO(5)$ which satisfy,
\begin{IEEEeqnarray}{ll}
[L_{ij}, L_{kl}] = i \left(\delta_{ik} L_{jl} + \delta_{jl} L_{ik} - \delta_{jk} L_{il} - \delta_{il} L_{jk} \right), \quad i,j,k,l = 1, \ldots, 5.
\end{IEEEeqnarray}
For the propagators of the easy scalars we have:
\begin{IEEEeqnarray}{ll}
{\color{Red}\langle \langle(\tilde{\varphi}_6)_{\bL\phantom{'}}^{\phantom{\dagger}} (\tilde{\varphi}_6)_{\bL'}^\dagger\rangle = \delta_{\bL, \bL'} \cdot K^{\ensuremath{\hat{m}_{\text{easy}}}^2}}, \qquad \hat{m}_{\text{easy}}^2 \equiv 2 L_1 L_2 + L_1 + 2 L_2. \label{PropagatorSO5easy}
\end{IEEEeqnarray}
\indent As in the case of the $SU(2)$ generators which were expressed in \eqref{t1FuzzySphericalHarmonics}--\eqref{t3FuzzySphericalHarmonics} by means of the $SU(2)$ fuzzy spherical harmonics, the $SO(5)$ $G$-matrices \eqref{G-matrices} can be expressed in terms of the $SO(5)$ spherical harmonics as follows: 
\begin{IEEEeqnarray}{ll}
G_1 = \sqrt{\frac{c_n}{10}} \cdot \left( \hat{Y}_{++} + \hat{Y}_{--} \right), \qquad & G_2 = -i \, \sqrt{\frac{c_n}{10}} \cdot \left( \hat{Y}_{++} - \hat{Y}_{--} \right), \label{G-FuzzySphericalHarmonics1} \\
G_3 = - \sqrt{\frac{c_n}{10}} \cdot \left( \hat{Y}_{-+} - \hat{Y}_{+-} \right), \qquad & G_4 = -i \, \sqrt{\frac{c_n}{10}} \cdot \left( \hat{Y}_{-+} + \hat{Y}_{+-} \right), \label{G-FuzzySphericalHarmonics2} \\
G_5 = - \sqrt{\frac{c_n}{5}} \cdot \hat{Y}_{00}, \label{G-FuzzySphericalHarmonics3}
\end{IEEEeqnarray}
where the constant $c_n$ has been defined in \eqref{ConstantSO5} above, and we have set 
\begin{IEEEeqnarray}{ll}
\hat{Y}_{\alpha\beta} \equiv \hat{Y}_{(\frac{1}{2}, \frac{1}{2}) \frac{1}{2} \frac{1}{2} \alpha \beta}, \qquad \hat{Y}_{0 0} \equiv \hat{Y}_{(\frac{1}{2}, \frac{1}{2}) 0 0 0 0}.
\end{IEEEeqnarray}
By using the relations \eqref{G-FuzzySphericalHarmonics1}--\eqref{G-FuzzySphericalHarmonics3} between the $G$-matrices and the fuzzy $SO(5)$ spherical harmonics, we compute the contraction,
\begin{align} 
\wick{\c1{\tilde{\varphi}_i} \cdot \tr[G_j \c1{\tilde{\varphi}_j}]} = K^{5/2}(x, y) \cdot G_{i}.
\end{align}
\paragraph{Scalar propagators in AdS} The bulk-to-bulk propagator of a massive scalar field in Euclidean AdS$_{d+1}$ is given by the following formula (see e.g.\ the set of lectures \cite{DHokerFreedman02}):
\begin{IEEEeqnarray}{c}
G_{\Delta}\left(x,z;y,w\right) = \frac{\Gamma\left(\Delta\right)\tilde{\eta}^{\Delta}}{2^{\Delta + 1}\pi^{d/2}\Gamma\left(\Delta - \frac{d}{2} + 1\right)} \cdot {_2}F_1\left(\frac{\Delta}{2},\frac{\Delta+1}{2},\Delta - \frac{d}{2} + 1,\tilde{\eta}^2\right), \qquad \label{ScalarPropagatorAdS1}
\end{IEEEeqnarray}
where $\Delta$ is the scaling dimension of the scalar field and $\tilde{m}$ its mass, while we have defined,
\begin{IEEEeqnarray}{c}
\tilde{\eta} \equiv \frac{2 z w}{z^2 + w^2 + \left(x - y\right)^2}, \qquad \tilde{m}^2 \equiv \Delta\left(\Delta - d\right) \equiv m^2 - \frac{d^2 - 1}{4}.
\end{IEEEeqnarray}
As before, the sets of coordinates $(x,z) = (x_0,\ldots,x_{d-2},z)$ and $(y,w) = (y_0,\ldots,y_{d-2},w)$ parametrize (Euclidean) AdS$_{d+1}$ in the Poincaré frame. We may obtain an alternative expression for the scalar propagator \eqref{ScalarPropagatorAdS1} by using the so-called quadratic transformation formulas for the hypergeometric function (see \cite{AbramowitzStegun65}, eq.\ 15.3.16). We find \cite{DHokerFreedman98b},
\begin{IEEEeqnarray}{l}
{_2}F_1\left(\frac{\Delta}{2},\frac{\Delta+1}{2},\Delta - \frac{d}{2} + 1,\tilde{\eta}^2\right) = \left(1 + \frac{1}{2s}\right)^{\nu + \frac{d}{2}} {_2}F_1\left(\nu + \frac{d}{2}, \nu + \frac{1}{2}, 2\nu + 1,-s^{-1}\right), \qquad \quad \label{QuadraticTransformation}
\end{IEEEeqnarray}
where we have defined,
\begin{IEEEeqnarray}{c}
s \equiv \frac{1 - \tilde{\eta}}{2\tilde{\eta}} = \frac{\left(z - w\right)^2 + \left(x - y\right)^2}{4 z w}, \qquad \nu \equiv \Delta - \frac{d}{2} = \sqrt{m^2 + \frac{1}{4}}.
\end{IEEEeqnarray}
The scalar propagator \eqref{QuadraticTransformation} can be further transformed by means of the linear transformation formulas of hypergeometric functions (see \cite{AbramowitzStegun65}, eq.\ 15.3.3). We get,
\begin{IEEEeqnarray}{l}
{_2}F_1\left(\frac{\Delta}{2},\frac{\Delta+1}{2},\Delta - \frac{d}{2} + 1,\tilde{\eta}^2\right) = \nonumber \\
\hspace{2cm}= \left(1 + \frac{1}{s}\right)^{\frac{1-d}{2}} \left(1 + \frac{1}{2s}\right)^{\nu + \frac{d}{2}} {_2}F_1\left(\nu - \frac{d}{2} + 1, \nu + \frac{1}{2}, 2\nu + 1,-s^{-1}\right), \qquad \label{LinearTransformation}
\end{IEEEeqnarray}
so that by plugging \eqref{LinearTransformation} into the expression for the scalar propagator \eqref{ScalarPropagatorAdS1}, we get:
\begin{IEEEeqnarray}{l}
G_{\Delta}\left(x,z;y,w\right) = \frac{1}{2^d \pi^{\frac{d+1}{2}}} \frac{\Gamma\left(\nu + \frac{d}{2}\right)\Gamma\left(\nu + \frac{3}{2}\right)}{\Gamma\left(2\nu + 2\right)} \cdot \frac{{_2}F_1\left(\nu - \frac{d}{2} + 1, \nu + \frac{1}{2}, 2\nu + 1,-s^{-1}\right)}{\left(1 + s\right)^{\frac{d-1}{2}} s^{\nu + \frac{1}{2}}}, \qquad \quad \label{ScalarPropagatorAdS2}
\end{IEEEeqnarray}
by also using the Legendre duplication formula for the gamma function, 
\begin{IEEEeqnarray}{l}
\Gamma\left(\nu + 1\right) = \frac{\sqrt{\pi}}{2^{2\nu + 1}} \cdot \frac{\Gamma\left(2\nu + 2\right)}{\Gamma\left(\nu + \frac{3}{2}\right)}.
\end{IEEEeqnarray}
\indent As it turns out \cite{NagasakiTanidaYamaguchi11, Buhl-MortensenLeeuwIpsenKristjansenWilhelm16c}, the propagators of the various fields in the (codimension-1) defect CFTs we are examining are related to the corresponding propagators in AdS, where the holographic directions $z$ and $w$ are replaced by the coordinates which are normal to the codimension-1 boundary. In our case the scalar fields propagate inside AdS$_4$, so that we set $d = 3$ in the above formulas for the propagators, and $(z,w) \rightarrow (x_3,y_3)$ for a boundary that is located at $x_3 = y_3 = 0$. By defining,
\begin{IEEEeqnarray}{l}
K^{m^2}\left(x,y\right) \equiv \frac{\gym^2}{2} \cdot \left(x_3 y_3\right)^{-\frac{d-1}{2}} \cdot G_{\Delta}\left(\left\{x_0,x_1,x_2\right\},x_3;\left\{y_0,y_1,y_2\right\},y_3\right),
\end{IEEEeqnarray}
we may use the expression \eqref{ScalarPropagatorAdS2} for the scalar propagator in AdS$_4$ to obtain,
\begin{IEEEeqnarray}{l}
K^{\nu}\left(x,y\right) = \frac{\gym^2}{16\pi^2}\frac{1}{{2\nu + 1 \choose \nu + \frac{1}{2}}}\frac{{}_2 F_1\left(\nu - \frac{1}{2}, \nu + \frac{1}{2}, 2\nu + 1 ; -\xi^{-1}\right)}{\left(1 + \xi\right)\xi^{\nu + \frac{1}{2}}}\cdot\frac{1}{x_3 y_3}, \qquad \label{ScalarPropagatorAdS3}
\end{IEEEeqnarray}
where $m$ is the mass of the scalar field and the dCFT invariant ratio $\xi$ was defined in \eqref{DefectInvariantRatios}:
\begin{IEEEeqnarray}{c}
\xi \equiv \frac{\left|x-y\right|^2}{4x_3y_3}, \qquad \nu \equiv \sqrt{m^2 +\frac{1}{4}},
\end{IEEEeqnarray}
for $x = \left(x_0,\ldots,x_3\right)$ and $y = \left(y_0,\ldots,y_3\right)$.

\bibliographystyle{JHEP}
\bibliography{bibliography}
\end{document}